\documentstyle[prl,aps,twocolumn]{revtex}

\begin{document}

\draft
\title{ Localized Structures Embedded in the Eigenfunctions of 
Chaotic Hamiltonian Systems }
\author{E. Vergini$^{a}$ and D.A. Wisniacki$^{b}$} 

\address{$^1$ Departamento de F\'{\i}sica, Comisi\'on Nacional de
Energ\'{\i}a At\'omica. 
 Av. del Libertador 8250, 1429 Buenos Aires, Argentina.}

\address{$^2$Departamento de F\'{\i}sica ``J.J. Giambiagi'', 
FCEN, UBA, Pabell\'on 1, Ciudad Universitaria, 1428 Buenos Aires, Argentina}

\date{\today}
\maketitle

\begin{abstract}
We study quantum localization phenomena in chaotic systems with a parameter.
The parametric motion of energy levels proceeds without crossing any other
and the defined avoided crossings quantify the interaction between states.
We propose the elimination of avoided crossings as the natural mechanism
to uncover localized structures.
We describe an efficient method for the
elimination of  avoided crossings in chaotic billiards and apply it
to the stadium billiard. We find many scars of short periodic orbits 
revealing the skeleton on which quantum mechanics is built. 
Moreover, we have
observed strong interaction between similar localized structures.

\end{abstract}

\centerline{PACS numbers: 05.45.+b, 03.65.Sq, 03.20.+i}

%
%

\narrowtext

The connection between individual eigenstates
and classical invariant sets in chaotic systems is far from being 
fully understood at present. 
A first step was given by Berry \cite{ber} and Voros
\cite{vor} twenty years ago. They conjectured that the eigenstates
are associated with the whole energy surface explored
ergodically by the orbits and then, the eigenstates would be locally 
like random superpositions of
plane waves. That picture is supported by Shnirelman's theorem \cite{shn}
and the pioneering numerical work by McDonald and Kaufman 
\cite{mcd} on the Bunimovich stadium billiard - an ergodic system 
\cite{bun}.
However, the paper by Heller \cite{hel} modified
that point of view. He emphasized that a large number of high excited
eigenfunctions for the stadium billiard has an enhancement along the
shorter periodic trajectories (scars). In 1988, Bogomolny \cite{bog}
developed the semiclassical theory of scars and explained the extra
density near unstable periodic trajectories. He represented the
probability density as a sum over a finite number of periodic orbits 
(see also \cite{ber2}). 

Another unexpected localization phenomenon was observed in the stadium billiard
for the first time; the famous bouncing ball ({\bf b.b}) states
\cite{mcd2}. They correspond to the family of neutral periodic orbits
living in the stadium. Localization is stronger than in the case
of scars and survives the semiclassical limit \cite{hel2}. The 
semiclassical theory of {\bf b.b} was developed recently by
Tanner \cite{tan}.

It is easy to observe {\bf b.b} states in the stadium billiard, 
but it is practically impossible to find complete series of
{\bf b.b} states. On the other hand, it is very difficult to
find a family of scars. That is, given a scar with
wave number $k$ which is associated to a particular
periodic orbit of lenght $\cal L$, we expect to see a family of
scars of the same orbit at $\;k+ 2 {\pi}n/{\cal L} \;$ \cite{hel}, with
$n$ an integer. But in general, a sequence of similar scars is
not observed. Such a phenomenon appears only for some 
short periodic orbits in some regions of the spectrum but it 
is impossible to define a systematic of scars \cite{sim}.

We believe that in order to study the mechanisms giving rise to
localization it is necessary first to clean the spectrum.
The individual properties of two well defined states depending on a
parameter are mixed when they collide in an avoided crossing 
\cite{zen}. Then, we expect that the elimination of avoided crossings 
({\bf a.c}) in systems governed by a parameter dependent Hamiltonian 
provides a mechanism for localization. But
in a chaotic system it is impossible to remove all {\bf a.c} because
an individual eigenstate interacts strongly with other states and the
interaction is of long range. However, 
we will extract only the {\bf a.c} with highest curvatures. 
Recently, it was stressed \cite{del}
that the distribution of highest curvatures has a universal character
and the tail of low curvatures is a particular property of the system
possibly associated with scarring phenomena. 
Precisely, {\it we want to remove
the universal property of chaotic eigenfunctions 
related to the Shnirelman's theorem,
and to retain fluctuations  revealed like phase space localization.}
On the same footing, Takami \cite{tak} has suggested that {\bf a.c} are 
contributed by long
periodic orbits. In the light of this, by removing {\bf a.c},
scars of short periodic orbits would arise. 

We have developed a simple method to 
remove {\bf a.c} in chaotic billiards governed by a shape parameter.
Let's $\phi_{\mu} ({\bf r} )$ and $\phi_{\nu} ({\bf r} )$ be normalized
eigenfunctions of a planar billiard with Dirichlet conditions on the
boundary $\cal C$. In ref. \cite{ver} the following
quasiorthogonality relation was shown 
\begin{equation}
\oint_{{\cal C}} \frac{\partial \phi_{\mu}}{\partial {\bf n}}
\;\frac{\partial \phi_{\nu}}{\partial {\bf n}}\;
\frac{r_{n}\;ds}{2 k_{\mu}  k_{\nu}}
= \delta_{\mu,\nu} + \frac{(k_{\mu}-k_{\nu})}{(k_{\mu}+k_{\nu})}
{\cal O}(1) \;\;\;,
\label{ort}
\end{equation}
where $k_{\mu}$ and $k_{\nu}$ are the corresponding eigenwave numbers,
$r_{n}\equiv {\bf r.n}$ with $\bf n$ the outward normal unit vector to
$\cal C$, and $s$ is arc lenght round $\cal C$.
Relation (\ref{ort}) defines an effective Hilbert space whose dimension
is of order $k$, and where an eigenstate is represented by 
$\;\varphi(s)\equiv \partial \phi / \partial {\bf n} ({\bf r} (s))\;$;
the eigenfunction in this space.

Now we deform the boundary by changing a shape parameter
$\ell$ ($\ell\!=\!{\ell }_{0}$ for $\cal C$). Using equation (\ref{ort})
it is possible to show \cite{vergi2} that  eigenenergies
($k^{2}$) and eigenfunctions at different values of the parameter
are connected
through the following parameter dependent Hamiltonian written in the
basis $\;\{ \varphi_{\mu} \}\;$ of eigenfunctions at ${\ell }_{0}$,
\begin{equation}
H_{\mu\nu}({\ell }_{0}+\delta \ell)\simeq k_{\mu}^{2}
\ \delta_{\mu,\nu}   + \; \delta \ell \; H'_{\mu\nu}\;\;,
\label{exp}
\end{equation}
with
\begin{equation}
H'_{\mu\nu}=-(\oint_{{\cal C}} r'_{n}\;\varphi_{\mu} \;\varphi_{\nu}\; ds)\times [1+(k_{\mu}\!-\!k_{\nu}){\cal O}(k^{-1})]
\label{def}
\end{equation}
where primes indicate derivation with respect to $\ell$. Relation (\ref{def})
was obtained previously \cite{ber3} for the case when 
$k_{\mu}\!=\!k_{\nu}$. By diagonalizing
$H(\ell_{1})$ we obtain the eigenenergies $\tilde k_{\mu}^{2}$
and the eigenfunctions $\tilde\varphi_{\mu}$ at $\ell_{1}$. Suppose that
at $\ell_{1}$ the states $\mu$ and $\nu$ collide in an {\bf a.c}.
In ref. \cite{maj} an efficient way of determining 
{\bf a.c} was described in terms of the coefficients 
$\;C_{\mu\nu}(\ell)\equiv <\varphi_{\mu}|\partial\varphi_{\nu}/\partial 
\ell>\;$
which define a driven evolution of the system. Using equation (\ref{exp}) 
and perturbation theory we obtain  
$\;C_{\mu\nu}(\ell)= H'_{\mu\nu}/(k_{\mu}^{2}-k_{\nu}^{2})\;$, where 
$\mu$ and $\nu$ are associated to eigenstates of $H(\ell)$ (the adiabatic
basis). Around $\ell_{1}$, 
$C_{\mu\nu}$ behaves like the Lorentzian function 
$\;(\ell_{int}/2)/[\ell_{int}^{2}+(\ell-\ell_{1})^{2}]\;$,
with $\;\ell_{int}=1/2 C_{\mu\nu}(\ell_{1})\;$.
The area in the energy spectrum, where the {\bf a.c} is relevant, can
be estimated by $\;A_{\mu\nu}=\Delta k^{2} \times \ell_{int}\;$. Then, 
if the area is lower than a prescribed value $v$
\begin{equation}
A_{\mu\nu}=(k_{\mu}^{2}-k_{\nu}^{2})^{2}/|H'_{\mu\nu}|<v\equiv
m\; \dot \ell\; \ln 2^{4}/\pi \hbar\;\;,
\label{cri}
\end{equation}
we eliminate the {\bf a.c} by transforming the diagonal matriz of 
$H(\ell_{1})$ as follows:
$\;\tilde k_{\mu}^{2}(new)=\tilde k_{\nu}^{2}(new)=
(\tilde k_{\mu}^{2}+\tilde k_{\nu}^{2})/2\;$. All {\bf a.c} which
satisfy (\ref{cri}) are eliminated in the increasing order of their area.
Criterion (\ref{cri}) has also a simple dynamic interpretation.
If the billiard contains a particle of mass $m$ and the boundary is driven
with a velocity $\dot \ell$, the {\bf a.c}
is eliminated when the Landau-Zener probability transition is greater than
one half.

The new Hamiltonian is less chaotic than the previous one,
and  the adiabatic basis of the new Hamiltonian is more adequate 
to study nonadiabatic effects \cite{lic}. For nearly integrable 
billiards, this new adiabatic basis goes to the diabatic one of the 
original Hamiltonian.
For chaotic billiards, the elimination of {\bf a.c} between nearest
neighboring levels gives rise to {\bf a.c} between distant levels 
(see Fig.~\ref{espe12}(a)).

We have studied the desymmetrized  stadium billiard with radius
$r$ and straight line of length $a$ \cite{maj}. The boundary only 
depends on the shape parameter $\ell = a/r$ (the area is fixed to the
value $\;1\!+\!\pi/4$).
Fig.~\ref{espectro} compares the approximated spectrum \cite{nota}
obtained from eq. (\ref{exp})
with the exact one; the agreement is excellent. We find structures
surviving parametric variation (straight lines interrupted by small
{\bf a.c}). The most evident being {\bf b.b} states, {\it e.g.} 
states 6, 14 and 18 of Fig.~\ref{fun}(a). But in general any localized 
eigenfunction has the same behavior.
In Fig.~\ref{espe12}(a) 
we show the spectrum of the transformed Hamiltonian by
elimination of all {\bf a.c} satisfying eq.(\ref{cri}), with 
$v=1.5$. At this stage all {\bf b.b} structures embedded in the
spectrum emerge clearly. {\bf b.b} states are identified by
the quantum numbers $(n_{x},n_{y})$ counting excitations
in the horizontal and vertical direction respectively;
$n_{y}$ labels {\bf b.b} series. The states 14, 18 and
23 of Fig.~\ref{fun}(b) correspond to $(n_{x},15)$, with 
$n_{x}=1,2\;{\rm and}\;3$ resp., and states 6, 16 and 28 to 
$(n_{x},14)$, with $n_{x}=6,7 \;{\rm and}\;8$.
At this stage also many scars of short periodic orbits clearly appear.
They are the states 1, 2, 4, 7, 8, 9, 12, 17, 20, 21, 22, 26, 28, 29 and 30 
of Fig.~\ref{fun}(b); 
the associated periodic orbits are identified in Fig.~\ref{orbi}. In Fig.~\ref{espe12}(b) we have eliminated
avoided crossings up to $v=5$ giving rise to new localized structures;
states 3, 5, 10, 11, 13, 24 and 27 of Fig.~\ref{fun}(b).
We have not identified  the
states 15 and 25 (probably contributed
by the whispering gallery family of periodic orbits) and the
state 19. 

The structures in Fig.~\ref{fun}(b) are obtained from the spectrum 
(Fig.~\ref{fun}(a)) by
an orthogonal transformation which reduces the parametric interaction among states; that is , the non-diagonal elements of the deformation matrix 
$H'$ are reduced considerably in the new basis.
The new states are characterized by the mean energy $<k^{2}>$ and the
dispersion $\sigma$ measured in units of the mean energy levels.
If $\sigma$ is lower than one, the new state has high probability
of appearing in the spectrum. For $\sigma\!=\!1$ the associated structure lives a time equal to the Heisenberg time. 
For scars of periodic trajectories we expect 
$\sigma$ to increase with $k$ according to the Shnirelman's theorem.
For {\bf b.b} states with fixed $n_{x}$, we expect
$\sigma$  to go to zero when $n_{y}$ go to infinite.
 
The non-diagonal elements of the original Hamiltonian in the new basis 
give us the interaction between localized structures. We have observed
strong interaction between similar structures. For example, {\bf b.b} 
states $(1,15)$ and 
$(2,15)$ have low interaction with the other states 
({\bf a.c} are very small -see Fig.~\ref{espectro}), except with the {\bf b.b}
state $(7,14)$ and with
the states 10 and 24 of Fig.~\ref{fun}(b).
Strong interaction was also observed between scars of short periodic orbits
which are close in phase space. For example, states 4 and 22 have strong
interaction with states 2 and 17 respectively, and states 10 and 24 with
states 11 and 27 respectively. In some situations, this strong interaction 
probably decides the occurrence of a defined structure. For example, 
states 4 and 22 are 
built principally with orbit ({\bf j}), but the Bohr-Sommerfeld 
quantization rule for this orbit ($k=0.6999\; (n +1/4)$) predicts
a scar at $k \sim 47.07$ which is not observed. However, the second
contribution, given by orbit ({\bf d}) ($k=1.3013\; (n +1/2)$),
does not predict such scar. In fact, in the range $35<k<55$ we have observed
structures like 4 and 22 only for those values of k which satisfy  
approximatelly quantization for ({\bf j}) and  ({\bf d}) simultaneously. 
A deep study
of this beating phenomenon and the interaction between periodic orbits, 
which appear very strong and can introduce
new insight into the semiclassical theory of chaotic systems, is 
presently under way \cite{vergi2}.

We transform the eigenfunctions in order to reduce the
parametric interaction between states. From a semiclassical point of view, 
the new structures are not contributed by long periodic orbits. The study of
the new states in terms of short periodic orbits is amenable to the
technique developed recently in ref. \cite{sim}.

Finally, we mention that the large parametric correlation observed by Tomsovic 
\cite{tom} in the stadium billiard (random matrix theory predicts no correlations)
is simply understood in terms of these localized 
structures. On the other hand, diffusion and dissipation 
in mesoscopic systems are affected by these localized structures and the
theory developped by Wilkinson \cite{wil2} based on random matrix theory 
needs to be revised.

We would like to thank M. Saraceno for useful suggestions. E.V. acknowledges
the hospitality of U. Smilansky and The Weizmann Institute of Science where
part of this work has been done.

\begin{figure}
\caption{The approximated spectrum (solid lines) obtained from eq. (2)
is compared with the exact spectrum (dots).}
\label{espectro}
\end{figure}

\begin{figure}
\caption{ (a) The spectrum of the tranformed Hamiltonian by elimination
of all avoided crossing satisfying eq. (\protect\ref{cri}) with 
$v=1.5$. The scarred states corresponding to Fig.\protect\ref{fun} (b) are shown.  
 (b) Idem (a) for $v=5$.}
\label{espe12}
\end{figure}

\begin{figure}
\caption{ (a) Linear density plots of the eigenfunctions for the 
stadium billiar with $\ell=1$ and area $1+\pi/4$. The numbers below
each plot are the label (left) and  the wave number $k$.
(b) Linear density plots of the eigenfunctions for the
transformed Hamiltonian by elimination
of all avoided crossing satisfying eq.(\protect\ref{cri}), with 
$v=5$. Letters on the top identify  periodic orbits
of Fig. \protect\ref{orbi} with higher contribution 
to each localized state.
The label, $\sqrt{<k^{2}>}$ and the dispersion $\sigma$ are showm 
below each plot. $\sigma$ is measured in units of the mean energy
levels.}
\label{fun}
\end{figure}

\begin{figure}
\caption{Several short periodic orbits of the desymmetrized stadium billiard
with  $\;\ell\! =\!1$.}
\label{orbi}
\end{figure}


\begin{references}
\bibitem{nota} We have multiplied the deformation matrix (eq. (3))
by the Lorentzian function $\;[1+Area (k_{\mu}^2-k_{\nu}^2)]^{-1}\;$ to
have into account the finite character of the effective Hilbert space.
\end{references}
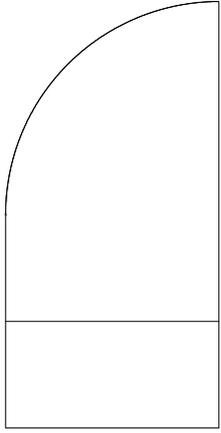
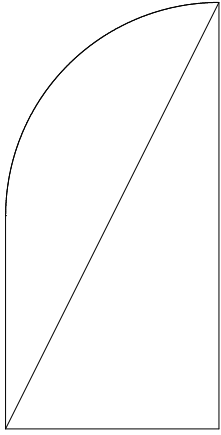
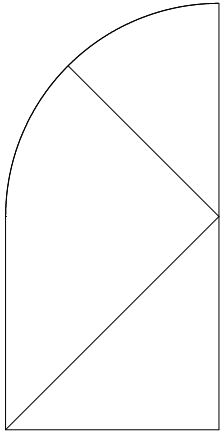
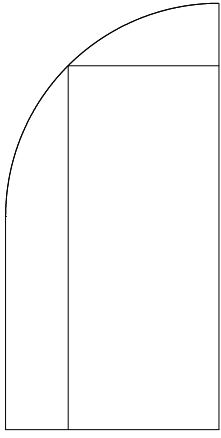
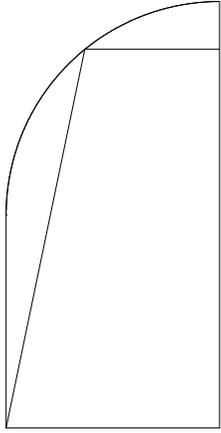
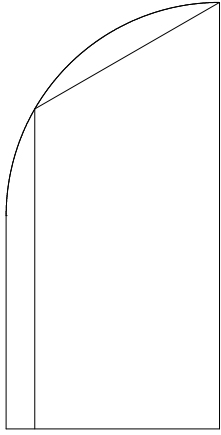
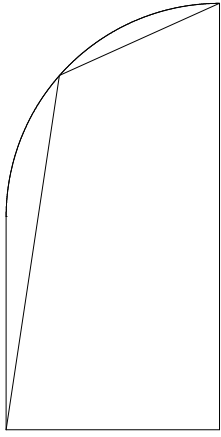
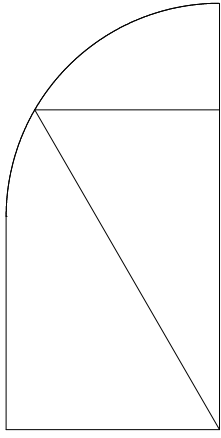
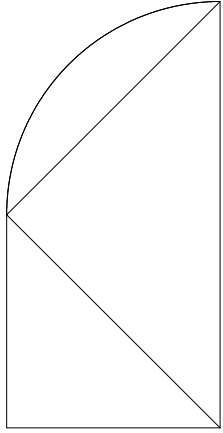
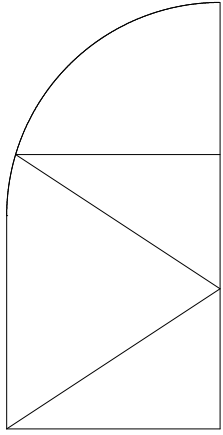
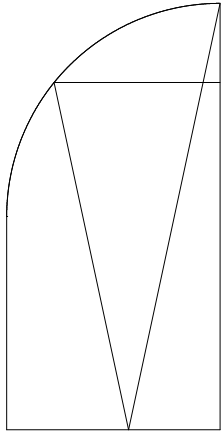
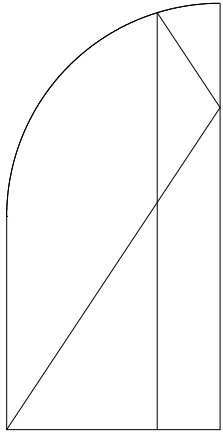
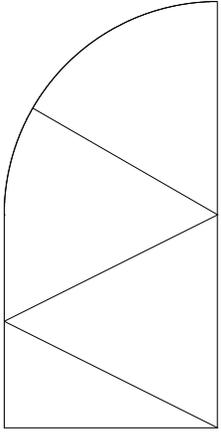
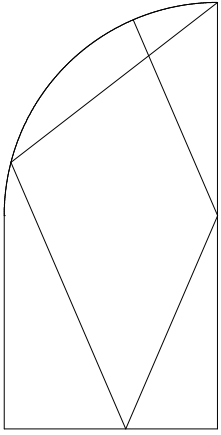
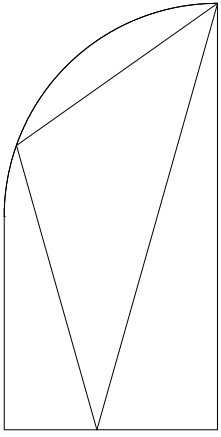
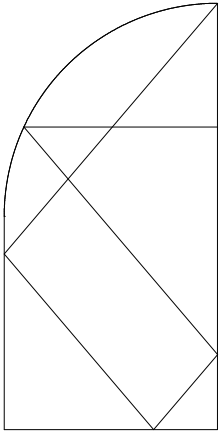
\end{document}